\documentclass[11pt,english]{article}
\usepackage[T1]{fontenc}
\usepackage[latin9]{inputenc}
\usepackage[a4paper]{geometry}
\usepackage{xcolor}
\usepackage{pdfcolmk}
\usepackage{amsmath}
\usepackage{amssymb}
\usepackage{setspace}
\PassOptionsToPackage{normalem}{ulem}
\usepackage{ulem}
\makeatletter

\providecolor{lyxadded}{rgb}{0,0,1}
\providecolor{lyxdeleted}{rgb}{1,0,0}

\newcommand{\lyxaddress}[1]{
\par {\raggedright #1
\vspace{1.4em}
\noindent\par}
}

\usepackage{babel}
\usepackage[unicode=true,
 bookmarks=false,
 breaklinks=false,
 pdfborder={0 0 1},
 backref=false,
 colorlinks=true]
 {hyperref}

\PassOptionsToPackage{normalem}{ulem}
\usepackage{ulem}
\let\cite@rig\cite
\newcommand{\b@xcite}[2][\%]{\def\def@pt{\%}\def\pas@pt{#1}
  \mbox{\ifx\def@pt\pas@pt\cite@rig{#2}\else\cite@rig[#1]{#2}\fi}}
\renewcommand{\underbar}[1]{{\let\cite\b@xcite\uline{#1}}}

\makeatother

\usepackage{babel}
\begin{document}

\title{Quaternion Generalization of Super Poincare Group }

\author{Bhupendra C. S. Chauhan and O. P. S. Negi}

\maketitle

\lyxaddress{\begin{center}
Department of Physics\\
Kumaun University\\
 S. S. J. Campus\\
Almora \textendash{} 263601 (Uttarakhand)\\
Email: bupendra.123@gmail.com \\
$\,\,\,\,$ops\_negi@yahoo.co.in
\par\end{center}}
\begin{abstract}
Super Poincare algebra in $D=6$ space-time dimensions has been analysed
in terms of quaternion analyticity of Lorentz group. Starting the
connection of quaternion Lorentz group with $SO(1,5)$ group, the
$SL(2,\mathbb{H})$ spinors for Dirac \& Weyl representations of Poincare
group are described consistently to extend the Poincare algebra to
Super Poincare algebra for $D=6$ space-time.
\end{abstract}

\section{Introduction }

The higher dimensional theories have become \cite{key-1,key-2} an
essential part for the modern development of self consistent field
theories. Since these are eligible for answering most of the hierarchy
anomalies that occurring at very high range of energy. Supersymmetry
and supergravity theories have a well and consistent structure, they
originates in a spontaneous way by the maximum extension of symmetries
of S-matrix of QFT\cite{key-3,key-4}. Higher dimensional Supersymmetric
theories \cite{key-5} are the most possible gauge theories in order
to understand the theories of everything (TOE). Previously, It has
been shown that supersymmetric theories are possible only for the
space-time dimensions of 3,4,6,10\cite{key-6} .Simultaneously the
connection between higher dimensional supersymmetric field theories
and division algebra has already been established by Kugo-Townsend\cite{key-7},
Lukereski-Topan\cite{key-8} and Seema-Negi\cite{key-9}. Likewise
the reduction of higher dimensional supersymmetric gauge theories
to lower dimensional space has also been studied explicitly by Schwartz-Brink\cite{key-10}.
On the other hand in view of Hurwitz theorem there exists \cite{key-11}
four normed division algebras $\mathbb{R}$,$\mathbb{C}$,$\mathbb{H}$
and$\mathcal{O}$ respectively named as the algebras of Real numbers,
Complex numbers, Quaternions and Octonions. It is pointed out that
by Kugo-Townsend \cite{key-7} that the supersymmetric gauge theories
are well examined for$D=3,4,6,10$ in terms of components of division
algebra respectively associated with the algebra of real numbers $\mathbb{R}$
(for D=3), of complex numbers $\mathbb{C}$ (for $D=4$), quaternions
$\mathbb{H}$ (for $D=6$) and octonions $\mathcal{O}$ $(D=10)$.

Keeping in view the utility of higher dimensional space time and the
physics beyond standard model focused on Supersymmetry, in the present
paper we have made an attempt to discuss Super Poincare algebra in
$D=6$ space-time dimensions by Quaternion algebra ($\mathbb{H}$).
The manuscript extensively studied the quaternion analyticity of Lorenz
group and its connection with $SO(1,5)$ group, the $SL(2,\mathbb{H})$
spinors, Dirac and Weyl representation of Poincare group followed
by the extension of Poincare algebra to Super Poincare algebra for
$D=6$ space-time.

\section{The Quaternion analyticity of Lorentz group:}

A proper Lorentz transformation in $D=4$ space is defined as 
\begin{align}
x'^{\mu}= & \Lambda_{\nu}^{\mu}x^{\nu}\label{eq:1}
\end{align}
which forms a non-compact Lie group $SO(1,3)$ satisfying the following
condition of metric $\quad$preserving group \cite{key-12} i.e.
\begin{align}
\Lambda^{-1}\eta\Lambda= & \eta\:\:\:\forall\Lambda\epsilon SO(1,3).\label{eq:2}
\end{align}
Here the metric is defined as ($\eta^{\mu\nu}=1,-1,-1,-1$). The Lorentz
group $SO(1,3)$ has universal covering group $SL(2,\mathbb{C})$
which is isomorphic to the projective group of M$\ddot{o}$bius transformation 

\begin{align}
f(z)= & \frac{az+b}{cz+d}\:\:\:\longleftrightarrow A=\left(\begin{array}{cc}
a & b\\
c & d
\end{array}\right)\:\:(\forall\, a,b,c,d\,\epsilon\mathbb{C}).\label{eq:3}
\end{align}
Let us construct the mapping from Minkowski space to the set of Hermitian
complex $2\times2$ Pauli matrices such that a four-vector is described
as 
\begin{align}
x^{\mu}\rightarrow\rho(x^{\mu})= & x^{\mu}\sigma_{\mu}=\left(\begin{array}{cc}
x^{0}-x^{3} & x^{1}+ix^{2}\\
x^{1}-ix^{2} & x^{0}+x^{3}
\end{array}\right)\label{eq:4}
\end{align}
where $\sigma_{0}=\hat{I}$, $\sigma_{j}$'s are $2\times2$ Pauli
spin matrices. In order to write the quaternion analysis of $SL(2,\mathbb{C})$
group, we define a quaternion \cite{key-13} as
\begin{align}
q= & q^{0}e_{0}+q^{j}e_{j}\:(\forall j=1,2,3)\label{eq:5}
\end{align}
where $(q^{0},q^{1},q^{2},q^{3}\epsilon R)$ and $e_{0}=\hat{1}$,
$e_{j}'$s are the quaternion units satisfying the following multiplication
rule 

\begin{align}
e_{j}e_{k}= & -\delta_{jk}+\epsilon_{jkl}e_{l}\:(\forall j,k,l=1\, to\,3).\label{eq:6}
\end{align}
Here $\delta_{jk}$ is the Kronecker delta symbol and $\epsilon_{jkl}$
is the three index Levi-Civita symbol. Quaternion units $e_{j}'s$
are well connected with Pauli matrices as $e_{1}\longleftrightarrow-i\sigma_{1},\, e_{2}\longleftrightarrow-i\sigma_{2},\, e_{3}\longleftrightarrow-i\sigma_{3}$.
The universal covering group of quaternions is $Sp(1,\mathbb{H})$.
Under the identification with Pauli matrices there is a correspondence
between $Sp(1,\mathbb{H}$) and $USp(2,\mathbb{C})$ \cite{key-14}
but they describe Euclidean transformation in $D=4$ space time (rotation
in $S^{3}$sphere).We may now generalized the $SL(2,\mathbb{C})$
group (i.e. the special linear group of $2\times2$ complex matrices)
to the SL(2,$\mathbb{H}$) group (i.e. the special linear group of
$2\times2$ quaternion matrices). Here we split one imaginary unit
i ($i\epsilon\mathbb{C}$) to the triplet $e_{j}\:(e_{j}\epsilon\mathbb{H}(\forall j=1,2,3))$.
So the Pauli matrices are generalized to 2x2 quaternion \cite{key-15}$\Gamma$
matrices as 
\begin{align}
\Gamma^{\mu}= & \left\{ \Gamma^{0},\Gamma^{1},\Gamma^{2},\Gamma^{3},\Gamma^{4},\Gamma^{5}\right\} \label{eq:7}
\end{align}
where 
\begin{align}
\Gamma^{0}= & \left(\begin{array}{cc}
1 & 0\\
0 & 1
\end{array}\right),\,\,\Gamma^{1}=\left(\begin{array}{cc}
0 & e_{1}\\
-e_{1} & 0
\end{array}\right),\,\,\Gamma^{2}=\left(\begin{array}{cc}
0 & e_{2}\\
-e_{2} & 0
\end{array}\right),\nonumber \\
\Gamma^{3}= & \left(\begin{array}{cc}
0 & e_{3}\\
-e_{3} & 0
\end{array}\right),\,\,\Gamma^{4}=\left(\begin{array}{cc}
0 & 1\\
1 & 0
\end{array}\right),\,\,\Gamma^{5}=\left(\begin{array}{cc}
1 & 0\\
0 & -1
\end{array}\right).\label{eq:8}
\end{align}
 As such the quaternion $2\times2$ $\Gamma$ matrices represent a$D=6$
space time. So, a six-dimensional Minkowski vector is thus defined
as
\begin{align}
x^{\mu}\rightarrow\rho(x^{\mu})=x_{\mu}\Gamma^{\mu}\Longrightarrow & \left(\begin{array}{cc}
x^{0}+x^{5} & x^{4}+e_{1}x^{1}+e_{2}x^{2}+e_{3}x^{3}\\
x^{4}-e_{1}x^{1}-e_{2}x^{2}-e_{3}x^{3} & x^{0}-x^{5}
\end{array}\right).\label{eq:9}
\end{align}
which has the determinant
\begin{align}
det(X)= & (x^{0})^{2}-(x^{1})^{2}-(x^{2})^{2}-(x^{3})^{2}-(x^{4})^{2}-(x^{5})^{2}=x^{\mu}x_{\mu}\label{eq:10}
\end{align}
followed by the metric $\eta^{\mu\nu}=(1,-1,-1,-1,-1,-1)$. For $SL(2,\mathbb{H})$
group, the determinant of eq. (\ref{eq:8}) turns out to be unity.
So, the M$\ddot{o}$bius transformation for $SL(2,\mathbb{H})$\cite{key-16}
are written as 
\begin{align}
f(q)= & \frac{qz+b}{qz+d}\:\:\:\longleftrightarrow A=\left(\begin{array}{cc}
a & b\\
c & d
\end{array}\right)\:\:(\forall a,b,c,d\,\epsilon\mathcal{\mathbb{H}}).\label{eq:11}
\end{align}
The matrix groups in quaternion case $S^{4}\rightarrow\mathbb{H}$
are the compact groups $Sp(2,\mathbb{H})$ and its subgroup $SU(2)\times SU(2)\cong spin(4)$
is related to $rotation$ in the sphere and the compact $SL(2,\mathbb{H})$
describes general non-Euclidean transformations. Thus there exists
is Lie algebra isomerism $SL(2,\mathbb{H})\cong SU^{*}(4)\cong SO(5,1)$\cite{key-14,key-16}.
However, the group $Sp(2,\mathbb{H})$ is isomorphic to $USp(4,\mathbb{C})$
which is $10$ dimensional. So $Sp(2,\mathbb{H})$ is well connected
with $SL(2,\mathbb{H})$ group. Moreover, there is a correspondence
between projective groups and uni-modular groups respectively associated
with the division algebra of real numbers ($\mathbb{R}$), Complex
numbers($\mathbb{C}$) and the quaternions ($\mathbb{H}$) i.e.

\begin{align}
Sl(2,\mathbb{R})\sim & P(1,R)\nonumber \\
Sl(2,\mathbb{C})\sim & P(1,C)\nonumber \\
Sl(2,\mathbb{H})\sim & P(1,\mathbb{H}).\label{eq:12}
\end{align}
Under the transformation of $SL(2,\mathbb{H})$, $X$ transforms as

\begin{align}
X^{'}= & YXY^{\dagger}\: with\: Y\epsilon SL(2,\mathbb{H})\label{eq:13}
\end{align}

\section{The Quaternion Lorentz group of $SO(1,5)$:}

Considering the metric in $D=6$ space as $\eta_{\mu\nu}=\{1,-1,-1,-1,-1,-1\}$,
$R\epsilon O(1,5)$ satisfy the condition (\ref{eq:2}) of metric
preserving group by

\begin{align}
R^{-1}\eta R= & \eta\:\:\:(\forall R\epsilon O(1,5)\label{eq:14}
\end{align}
where $R$ is to be taken as 
\begin{alignat}{1}
R= & \left(\begin{array}{cccccc}
r_{00} & r_{01} & r_{02} & r_{03} & r_{04} & r_{05}\\
r_{10} & r_{11} & r_{12} & r_{13} & r_{14} & r_{15}\\
r_{20} & r_{21} & r_{22} & r_{23} & r_{24} & r_{25}\\
r_{30} & r_{31} & r_{32} & r_{33} & r_{34} & r_{35}\\
r_{40} & r_{41} & r_{42} & r_{43} & r_{44} & r_{45}\\
r_{50} & r_{51} & r_{52} & r_{53} & r_{54} & r_{55}
\end{array}\right)\label{eq:15}
\end{alignat}
Substituting this to eq.(14) we get $r_{00}=r_{11}=r_{22}=r_{33}=r_{44}=r_{55}=0,\: r_{ij}=-r_{ji},\: r_{oi}=r_{i0}\:(\forall i,j=1\, to\,5)$,
according the generators for$SO(1,5)$ group may be written. The determinant
of R turns out to be unity. Thus we may easily define the rotation
and Lorentz boost generators of $SO(1,5)$ group respectively denoted
by $L_{ij}$ and $N_{0i}$. For $SO(1,5)$ group there exists $10$
generators of rotation associated with $L_{ij}$ matrices followed
by $5$ generators corresponding to the Lorentz boost matrices $N_{0i}$.
Both $L_{ij}$ and $N_{0i}$ are traceless matrices. The five Lorentz
boost generators $N_{0i}$ are symmetric, while the other ten rotation
generators $L_{ij}$ are antisymmetric. Matrices $L_{ij}$ and $N_{0i}$
are discussed in the appendix-I.We have used the mappings $J_{ij}=iL_{ij}$
and $K_{oi}=iN_{0i}\,(\forall i,j=1,2\, to\,5)$. 

So the commutation relations are obtained as
\begin{align}
[J_{ij},J_{kl}]= & i(\delta_{ik}J_{jl}+\delta_{jl}J_{ik}-\delta_{jk}J_{il}-\delta_{il}J_{jk})\:\left\{ \forall i,j,k,l=1\, to\,5\right\} \nonumber \\
{}[K_{0i},K_{0j}]= & -iJ_{ij}\,\left\{ \forall i,j=1\, to\,5\right\} \nonumber \\
{}[K_{0i},J_{jk}]= & i(\delta_{ik}K_{0j}-\delta_{ij}K_{0k})\:\left\{ \forall i,j,k=1\, to\,5\right\} .\label{eq:16}
\end{align}
The generators $J_{ij}$ and $K_{0j}$ are respectively associated
with the generators of angular momentum and Lorentz boosts. These
equations can be combined together in tensorial form by taking $\mbox{\ensuremath{M}}_{ij}=J_{ij}$
and $M_{0i}=K_{0i}\,(\forall i,j=1,2\, to\,5)$. So the algebra of
quaternion Lorentz group $SO(1,5)$ describes the following structure
\begin{align}
[M_{\mu\nu},M_{\rho\sigma}]=- & i(\eta_{\mu\rho}M_{\nu\sigma}+\eta_{\nu\sigma}M_{\nu\rho}-\eta_{\mu\sigma}M_{\nu\rho}-\eta_{\nu\rho}M_{\mu\sigma})\label{eq:17}
\end{align}
where the metric for $SO(1,5)$ group is defined as$\{\eta_{\mu\nu}=+1,-1,-1,-1,-1,-1\}$.

\section{Formulation of $SL(2,\mathbb{H})$}

Let us define a vector $x^{\mu}\,\{\forall\mu=0,1,2,3,4,5\}$ in $D=6$
space as 

\begin{align}
x^{\mu}= & (x^{0},x^{1},x^{2},x^{3},x^{4},x^{5})=(x^{0},\vec{x})\label{eq:18}\\
x_{\mu}= & \eta_{\mu\nu}x^{\nu}=(x^{0},-x^{1},-x^{2},-x^{3},-x^{4},-x^{5})=(x^{0},-\vec{x})\label{eq:19}
\end{align}
 Likewise, the six $\Gamma$- matrices of eq.(\ref{eq:8}) can be
described in contra and covariant matrices as

\begin{align}
\Gamma^{\mu}= & (\Gamma^{0},\Gamma^{i}),\,\;\;(i=1\, to\,5)\label{eq:20}\\
\Gamma_{\mu}= & \eta_{\mu\nu}\Gamma^{\mu}=\tilde{\Gamma}^{\mu}=(\Gamma^{0},-\Gamma^{i})\,\:\:(i=1\, to\,5)\label{eq:21}
\end{align}
The $\Gamma$ matrices thus satisfy the Clifford algebra relation

\begin{align}
\Gamma^{\mu}\tilde{\Gamma}^{\nu}+\Gamma^{\nu}\tilde{\Gamma}^{\mu}= & 2\eta^{\mu\nu}\label{eq:22}
\end{align}
Similarly $Tr(\Gamma^{\mu}\tilde{\Gamma}^{\nu})=2\eta^{\mu\nu}$ where
the trace is defined as $Tr(P)=Re[Tr(P)]$ for $P\epsilon SL(2,\mathbb{H})$.
It is to be noted that the $\Gamma^{\mu}$ matrices are quaternion
Hermitian matrices $\Gamma^{\mu\dagger}=\Gamma^{\mu}$ where adjoint
of matrices is for the transpose of quaternion conjugation operation\cite{key-15}.
The Lorentz group$SO(1,5)$ in $D=6$ space-time is homomorphic to
the $SL(2,\mathbb{H})$group of 2x2 quaternion matrices i.e.
\begin{align}
h= & \left(\begin{array}{cc}
a & b\\
c & d
\end{array}\right),\,(\forall h\epsilon SL(2,\mathbb{H})\:\: det\, h=1)\label{eq:23}
\end{align}
where $a,b,c,d$ are quaternion numbers. Under the Lorentz transformation
the invariant quantity is related to the X as 
\begin{align}
det\, X= & \eta_{\mu\nu}x^{\mu}x^{\nu},\label{eq:24}
\end{align}
which is also invariant under the transformation of $SL(2,\mathbb{H})$
group. Since any quaternion matrix can be written as linear combination
of $\Gamma^{\mu}$ matrices, we write the elements 'h' of the group
$SL(2,\mathbb{H})$ as 
\begin{align}
h= & h_{\mu}\Gamma^{\mu}\label{eq:25}
\end{align}
where $h_{0},h_{k}(k=1\, to\,5)$ are quaternion numbers and $\Gamma^{\mu}$
matrices are defined in eq.(\ref{eq:8}). The change in $x^{\mu}$
under the Lorentz transformation can be described as
\begin{align}
x'^{\mu}= & \Lambda_{\nu}^{\mu}x^{\nu}\label{eq:26}
\end{align}
where $\Lambda_{\nu}^{\mu}$ is an element of Lorentz group in $D=6$
space-time. The corresponding transformation of $X$ that leaves invariant
the $detX=\eta_{\mu\nu}x^{\mu}x^{\nu}$ , is described under the $SL(2,\mathbb{H})$
group as

\begin{align}
X'= & hXh^{\dagger}\label{eq:27}
\end{align}
where $h^{\dagger}$ is quaternion conjugate of h. Since the scalar
product $\eta_{\mu\nu}x^{\mu}x^{\nu}$ is invariant under Lorentz
transformation, we should have\cite{key-17}

\begin{align}
det\, X= & det\, X'.\label{eq:28}
\end{align}
using the properties $Tr(\Gamma^{\mu}\tilde{\Gamma}^{\nu})=2\eta^{\mu\nu}$
and $Tr(\tilde{\Gamma}^{\mu}\Gamma_{\nu})=2\delta_{\nu}^{\mu}$ we
get
\begin{align}
x'^{\mu}= & \frac{1}{2}Tr(\tilde{\Gamma}^{\mu}hXh^{\dagger})=\frac{1}{2}Tr(\tilde{\Gamma}^{\mu}h\Gamma_{\nu}h^{\dagger})x^{\nu}.\label{eq:29}
\end{align}
This equation gives the explicit relation between $\Lambda_{\nu}^{\mu}$
elements of Lorentz group in $D=6$ space to the elements of SL(2,$\mathbb{H}$)
group. The group homomorphism between$\Lambda_{\nu}^{\mu}$ and $SL(2,\mathbb{H})$
may then be described by the relation between the components of $\Lambda_{\nu}^{\mu}$
and $h(h\epsilon SL(2,\mathbb{H}))$as 
\begin{align}
\Lambda_{0}^{0}= & \left|h_{0}\right|^{2}+\sum_{k=1}^{5}\left|h_{k}\right|^{2}\nonumber \\
\Lambda_{k}^{0}= & Re[(h_{0}\overline{h}_{k}+h_{k}\overline{h_{0}})+\epsilon_{ijk}(h_{0}e_{j}\overline{g_{i}}+h_{j}e_{i}\overline{h_{0}})+h_{5}e_{k}\overline{h_{4}}-h_{4}e_{k}\overline{h_{5}}]\nonumber \\
\Lambda_{4}^{0}= & Re[(h_{0}\overline{h_{4}}+h_{4}\overline{h_{0}})+\sum_{k=1}^{3}(h_{k}e_{k})\overline{h_{5}}-h_{5}\sum_{k=1}^{3}(e_{k}\overline{h_{k}})]\nonumber \\
\Lambda_{5}^{0}= & Re[(h_{0}\overline{h_{5}}+h_{5}\overline{h_{0}})+h_{4}\sum_{k=1}^{3}(e_{k}\overline{h_{k}})-\sum_{k=1}^{3}(h_{k}e_{k})\overline{h_{4}}]\nonumber \\
\Lambda_{0}^{k}= & Re[e_{k}(h_{4}\overline{h_{5}}-h_{5}\overline{h_{4}})-e_{k}\sum_{i=1(i\neq k)}^{3}(h_{i}e_{i}\overline{h_{0}}+h_{0}e_{i}\overline{h_{i}})-e_{i}(h_{i}e_{i}\overline{h_{0}}+h_{0}e_{i}\overline{h_{i}})]\nonumber \\
\Lambda_{0}^{4}= & Re[(h_{0}\overline{h_{4}}+h_{4}\overline{h_{0}})+h_{5}\sum_{i=1}^{3}e_{i}\overline{h_{i}}-\sum_{j=1}^{3}h_{j}e_{j}\overline{h_{5}}]\nonumber \\
\Lambda_{0}^{5}= & Re[(h_{0}\overline{h_{5}}+h_{5}\overline{h_{0}})-h_{4}\sum_{i=1}^{3}e_{i}\overline{h_{i}}+\sum_{j=1}^{3}h_{j}e_{j}\overline{h_{4}}]\label{eq:30}
\end{align}
where $\epsilon_{ijk}$ is Levi-Civita tensor and $i,j,k\rightarrow1,2,3$
and $\bar{h}_{i}$ is the quaternion conjugate of $h_{i}$. So, the
homomorphism between quaternion Lorentz group and SL(2,$\mathbb{H}$)
group is established in terms of the following properties i.e. 

(i) $\Lambda_{0}^{0}\geq1$ since the norms of division algebras $\mathbb{\mathbb{R}},\mathbb{\mathbb{C}},\mathbb{H}$,$\mathcal{O}$
is always positive . 

(ii) $\Lambda_{\nu}^{\mu}(M_{1})\Lambda_{\rho}^{\nu}(M_{2})=\Lambda_{\rho}^{\mu}(M_{1}M_{2})\,\forall\, M_{1},M_{2}\epsilon SL(2,\mathbb{H})$
can be easily verified by the cyclic property of the trace of Quaternion
matrices.

It is obvious that $\Lambda_{\nu}^{\mu}(M^{-1})=(\Lambda_{\nu}^{\mu}(M))^{-1}.$
The $SL(2,\mathbb{H})$ group is compact group hence the homomorphic
mapping of $\Lambda_{\nu}^{\mu}$ is continuous into $SL(2,\mathbb{H})$.
So by $det\,(\Lambda_{\nu}^{\mu}(1_{SL(2,\mathbb{H})}))=1$ it can
be stated obviously that $det(\Lambda_{\nu}^{\mu})=1$. Hence the
homomorphism between $SL(2,\mathbb{H})$ group of quaternion matrices
and Lorentz group in $D=6$ space has been established consistently.

\section{Quaternion Spinors in $SL(2,\mathbb{H})$:}

Let us write a vector and its conjugate in $SL(2,\mathbb{H})$ representation
as
\begin{align}
X= & x_{\mu}\Gamma^{\mu}=\left(\begin{array}{cc}
x^{0}-x^{5} & -x^{4}-e_{1}x^{1}-e_{2}x^{2}-e_{3}x^{3}\\
-x^{4}+e_{1}x^{1}+e_{2}x^{2}+e_{3}x^{3} & x^{0}+x^{5}
\end{array}\right)=\left(\begin{array}{cc}
x^{0}-x^{5} & x^{q}\\
x^{q\dagger} & x^{0}+x^{5}
\end{array}\right)\label{eq:31}
\end{align}

\begin{align}
\bar{X}= & x_{\mu}\tilde{\Gamma}^{\mu}=\left(\begin{array}{cc}
x^{0}+x^{5} & x^{4}+e_{1}x^{1}+e_{2}x^{2}+e_{3}x^{3}\\
x^{4}-e_{1}x^{1}-e_{2}x^{2}-e_{3}x^{3} & x^{0}-x^{5}
\end{array}\right)=\left(\begin{array}{cc}
x^{0}+x^{5} & -x^{q}\\
-x^{q\dagger} & x^{0}-x^{5}
\end{array}\right)\label{eq:32}
\end{align}
where $x^{q}$ is a quaternion element and $x^{q\dagger}$ is quaternion
conjugate of $x^{q}$ . The $X$ of equation (\ref{eq:31}) transforms
as vector under endomorphic transformation in $SL(2,\mathbb{H})$,
while $\bar{X}$ can be obtained from $X$ by space inversion operation
. Similarly the two component quaternion spinors are defined as

\begin{align}
\Psi_{\alpha}= & \left(\begin{array}{c}
\phi_{\alpha}\\
\chi_{\alpha}
\end{array}\right),\;\;\;(\forall\phi_{\alpha},\chi_{\alpha}\epsilon\mathbb{H})\nonumber \\
\Psi_{\alpha}^{\dagger}= & \left(\begin{array}{cc}
\phi_{\alpha}^{\dagger} & \chi_{\alpha}^{\dagger}\end{array}\right)\label{eq:33}
\end{align}
where $'\dagger'$ correspond to the quaternion Hermitian conjugate
operation on$\Psi$. The transformation properties under $SL(2,\mathbb{H})$
of undotted spinor and it's conjugate are such as

\begin{align}
\Psi'_{\alpha}= & M_{\alpha}^{\beta}\Psi_{\beta}\quad\quad\Psi_{\alpha}^{\dagger'}=\Psi_{\beta}^{\dagger}M_{\alpha}^{\beta\dagger}\quad M\epsilon SL(2,\mathbb{H})\label{eq:34}
\end{align}
while the dotted spinors and its conjugate transform\cite{key-12}
as

\begin{align}
\eta^{\overset{.}{\alpha}}= & \left(\begin{array}{c}
\zeta^{\overset{.}{\alpha}}\\
\xi^{\overset{.}{\alpha}}
\end{array}\right),\quad\:(\forall\zeta^{\overset{.}{\alpha}},\xi^{\overset{.}{\alpha}}\epsilon\mathbb{H})\nonumber \\
\eta^{\overset{.}{\alpha\dagger}}= & \left(\begin{array}{cc}
\zeta^{\overset{.}{\alpha}\dagger} & \xi^{\overset{.}{\alpha}\dagger}\end{array}\right)\label{eq:35}
\end{align}
where transformation properties under $SL(2,\mathbb{H})$ :

\begin{align}
\eta^{\overset{.}{\alpha}}= & \left(M^{-1\dagger}\right)_{\overset{.}{\beta}}^{\overset{.}{\alpha}}\eta^{\overset{.}{\beta}}\quad\;\eta^{\overset{.}{\alpha\dagger}}=\eta^{\overset{.}{\beta\dagger}}\left(M^{-1}\right)_{\overset{.}{\beta}}^{\overset{.}{\alpha}}\label{eq:36}
\end{align}
So the differential operator is defined as

\begin{align}
\partial=\Gamma^{\mu}\partial_{\mu}= & \left(\begin{array}{cc}
\partial^{0}-\partial^{5} & -\partial^{4}-e_{1}\partial^{1}-e_{2}\partial^{2}-e_{3}\partial^{3}\\
-\partial^{4}+e_{1}\partial^{1}+e_{2}\partial^{2}+e_{3}\partial^{3} & \partial^{0}+\partial^{5}
\end{array}\right)\label{eq:37}
\end{align}
which acts on a two component quaternion spinor.

\section{$8\times8$ Dirac Representation and Quaternion realization of Poincare
group in $D=6$ Space:}

The pure Lorentz boost transformation does not form closed group.
However it is embedded with the group of rotation generators $J_{i}'s$
in general Lorentz group. But both of these behave differently under
parity transformation $J_{i}\rightarrow J_{i}$ and $K_{i}\rightarrow-K_{i}$\cite{key-18}.
So there are two different spinor representations for spin 1/2 particles
in Lorentz group. Both spinor representation transform differently
under Lorentz transformations and parity. These are called left handed
$\left(\dfrac{1}{2},0\right)$and right handed $\left(0,\dfrac{1}{2}\right)$
spinors. So for a theory where parity conservation is required Dirac
representations are used conveniently over other representations,
because it corresponds to the representation of direct sum of $\left(\dfrac{1}{2},0\right)\oplus\left(0,\dfrac{1}{2}\right)$
spinors\cite{key-12}. 

It is stated earlier that theories which transform as a linear representation
of supersymmetry must have same number of bosonic and fermionic degrees
of freedom. So for $D=6$ dimensions the massless vector particle
acquires $D-2=6-2=4$ degrees of freedom. While a spinor is described
by $2^{D/2}=2^{3}=8$ dimensions\cite{key-10}. Therefore for the
generalization of the theory to supersymmetric case we need 8 dimensional
$\Upsilon-$ matrices. Hence we adopt the procedure to extend from
$2\rightarrow4$ and $4\rightarrow8$ dimensions. Let us write the
4-dimensional generalization of $\Gamma$ matrices are described in
terms of following 4x4 Dirac matrices to update quaternion supersymmetrization
i.e. 
\begin{align}
\gamma_{D}^{\mu}= & \left\{ \left(\begin{array}{cc}
\Gamma^{0} & 0\\
0 & -\Gamma^{0}
\end{array}\right),\left(\begin{array}{cc}
0 & \Gamma^{1}\\
-\Gamma^{1} & 0
\end{array}\right),\left(\begin{array}{cc}
0 & \Gamma^{2}\\
-\Gamma^{2} & 0
\end{array}\right),\left(\begin{array}{cc}
0 & \Gamma^{3}\\
-\Gamma^{3} & 0
\end{array}\right),\left(\begin{array}{cc}
0 & \Gamma^{4}\\
-\Gamma^{4} & 0
\end{array}\right),\left(\begin{array}{cc}
0 & \Gamma^{5}\\
-\Gamma^{5} & 0
\end{array}\right)\right\} \label{eq:38}
\end{align}
This representation of $\gamma-$ matrices satisfies the Clifford
algebra relation 
\begin{align}
\gamma_{D}^{\mu}\gamma_{D}^{\nu}+\gamma_{D}^{\nu}\gamma_{D}^{\mu}= & 2\eta^{\mu\nu}\label{eq:39}
\end{align}
Accordingly, the $\gamma_{D}^{6}$ matrix is described as
\begin{align}
\gamma_{D}^{6}= & \gamma_{D}^{0}\gamma_{D}^{1}\gamma_{D}^{2}\gamma_{D}^{3}\gamma_{D}^{4}\gamma_{D}^{5}\label{eq:40}
\end{align}
which comes out to be
\begin{align}
\gamma_{D}^{6}= & \left(\begin{array}{cc}
0 & I\\
I & 0
\end{array}\right).\label{eq:41}
\end{align}
Here $\gamma_{D}^{6}$ has the same properties as $\gamma_{5}$ do
in $D=4$ space. It decides the nature of the currents $\bar{\psi}\gamma_{D}^{6}\psi$
and $\bar{\psi}\gamma_{D}^{\mu}\gamma_{D}^{6}\psi$ under the Lorentz
transformation in $D=6$ space, which comes out to pseudoscalar and
pseudovector respectively. The generator of Lorentz transformation
in $4\times4$ matrix representation are now be described as 
\begin{align}
\Sigma_{D}^{\mu\nu}= & \frac{i}{4}(\gamma_{D}^{\mu}\gamma_{D}^{\nu}-\gamma_{D}^{\nu}\gamma_{D}^{\mu})\label{eq:42}
\end{align}
which satisfy the commutation rules of the Lie algebra of Lorentz
group in $D=6$ space as
\begin{align}
[\Sigma^{\mu\nu},\Sigma^{\rho\sigma}]=-i & (\eta^{\mu\rho}\Sigma^{\nu\sigma}+\eta^{\nu\sigma}\Sigma^{\mu\rho}-\eta^{\mu\sigma}\Sigma^{\nu\rho}-\eta^{\nu\rho}\Sigma^{\mu\sigma})\label{eq:43}
\end{align}
along with
\begin{align}
[\Sigma^{\mu\nu},\gamma_{D}^{\rho}]= & i(\eta^{\nu\rho}\gamma_{D}^{\mu}-\eta^{\mu\rho}\gamma_{D}^{\nu})\label{eq:44}
\end{align}
The $8\times8$ fully reducible quaternion generalization of$\gamma-$
matrices is now be taken as 
\begin{align}
\Upsilon_{D}^{\mu}= & \left\{ \left(\begin{array}{cc}
\gamma_{D}^{0} & 0\\
0 & \gamma_{D}^{0}
\end{array}\right),\left(\begin{array}{cc}
\gamma_{D}^{1} & 0\\
0 & \gamma_{D}^{1}
\end{array}\right),\left(\begin{array}{cc}
\gamma_{D}^{2} & 0\\
0 & \gamma_{D}^{2}
\end{array}\right),\left(\begin{array}{cc}
\gamma_{D}^{3} & 0\\
0 & \gamma_{D}^{3}
\end{array}\right),\left(\begin{array}{cc}
\gamma_{D}^{4} & 0\\
0 & \gamma_{D}^{4}
\end{array}\right),\left(\begin{array}{cc}
\gamma_{D}^{5} & 0\\
0 & \gamma_{D}^{5}
\end{array}\right)\right\} \label{eq:45}
\end{align}
and the $8\times8$ Lorentz generators are defined as 
\begin{align}
\Xi^{\mu\nu}=\dfrac{i}{4}( & \Upsilon_{D}^{\mu}\Upsilon_{D}^{\nu}-\Upsilon_{D}^{\nu}\Upsilon_{D}^{\mu}).\label{eq:46}
\end{align}
This equation satisfies the commutation relation of Lorentz group
in $D=6$ space-time as
\begin{align}
[\Xi^{\mu\nu},\Xi^{\rho\sigma}]=-i & (\eta^{\mu\rho}\Xi^{\nu\sigma}+\eta^{\nu\sigma}\Xi^{\mu\rho}-\eta^{\mu\sigma}\Xi^{\nu\rho}-\eta^{\nu\rho}\Xi^{\mu\sigma}).\label{eq:47}
\end{align}
Thus, we get the remaining commutation relations of Poincare algebra
in $D=6$ space as
\begin{align}
[P^{\mu},P^{\nu}]= & 0\nonumber \\
{}[\Xi^{\mu\nu},P^{\rho}]= & i(\eta^{\nu\rho}P{}^{\mu}-\eta^{\mu\rho}P^{\nu})\nonumber \\
{}[\Xi^{\mu\nu},\Xi^{\rho\sigma}]= & -i(\eta^{\mu\rho}\Xi^{\nu\sigma}+\eta^{\nu\sigma}\Xi^{\mu\rho}-\eta^{\mu\sigma}\Xi^{\nu\rho}-\eta^{\nu\rho}\Xi^{\mu\sigma})\label{eq:48}
\end{align}

\section{Weyl Basis for Chiral Representation : }

However, it is well known that at extreme relativistic limit the fermions
behave differently to electroweak interaction (called helicity conserved
interactions). Yet the Lagrangian of standard model doesn't remain
parity conserved. So, the Weyl representation is advantageous because
it separates the left handed spinors to right handed spinors\cite{key-19,key-20}.
As such, the 4-dimensional generalization of $\Gamma-$matrices given
by equation (\ref{eq:8}) is generalized in terms of following $4\times4$
Weyl matrices to update quaternion supersymmetrization i.e. 

\begin{align}
\gamma_{W}^{\mu}= & \left(\begin{array}{cc}
0 & \Gamma^{\mu}\\
\tilde{\Gamma}^{\mu} & 0
\end{array}\right)\label{eq:49}
\end{align}
 Likewise, the properties of $\gamma_{W}^{\mu}-$ matrices also satisfy
the Clifford algebra relation (\ref{eq:39}). As such, equation (\ref{eq:49})
leads to the conclusion that the $\gamma_{W}^{0}$ matrices in Weyl
representation is non-diagonal. On the other hand, the diagonal representation
of Weyl matrices is associated with the pseudoscalar matrix as:

\begin{align}
\gamma_{W}^{6}= & \gamma_{W}^{0}\gamma_{W}^{1}\gamma_{W}^{2}\gamma_{W}^{3}\gamma_{W}^{4}\gamma_{W}^{5}=\left(\begin{array}{cc}
-1 & 0\\
0 & 1
\end{array}\right)\label{eq:50}
\end{align}
which comes out to be diagonal stating that the solutions of Weyl
equations are the eigenstates of helicity rather than energy (contrary
of Dirac representation). Thus the $\gamma-$ matrices together with$\gamma_{W}^{6}$
satisfy the following properties

(i) $\gamma_{W}^{0}=\gamma_{W}^{0\dagger}$, 

(ii) $\gamma_{W}^{i}=-\gamma_{W}^{i\dagger}\;\{i=1,2,3,4,5\}$, 

(iii) $\gamma_{W}^{6}=\gamma_{W}^{6\dagger}$. 

Similarly, we get $\gamma_{W}^{0}\gamma_{W}^{\mu}\gamma_{W}^{0}=\gamma_{W}^{\mu\dagger}$.
Here $\gamma_{W}^{6}$ behaves in the same way as $\gamma_{5}$ do
in $D=4$ space. The projection operator for massless fermion $\frac{(1-\gamma_{W}^{6})}{2}$
associated with the left handed particles and while left handed antiparticles
are associated with $\frac{(1+\gamma_{W}^{6})}{2}$ . So, it acts
as chiral operator for massless fermions. However, the eigenvalue
of $\gamma_{W}^{6}$ is same as that of the helicity operator for
particle while it goes reversed to that of helicity operator for antiparticles.
Thus chirality and helicity has same meaning for particles but they
have opposite nature for antiparticles. Hence, the generators of homogeneous
Lorentz group associated with angular momentum and Lorentz boosts
$\{\Sigma^{\mu\nu}\:(\forall\mu,\nu=0\, to\,5)\}$ in $D=6$ space
are described as 4x4 matrices in the following manner i.e.
\begin{align}
\Sigma^{\mu\nu} & =\frac{i}{4}(\gamma_{W}^{\mu}\gamma_{W}^{\nu}-\gamma_{W}^{\nu}\gamma_{W}^{\mu})=\frac{i}{4}\left(\begin{array}{cc}
\Gamma^{\mu}\tilde{\Gamma}^{\nu}-\Gamma^{\nu}\tilde{\Gamma}^{\mu} & 0\\
0 & \tilde{\Gamma}^{\mu}\Gamma^{\nu}-\tilde{\Gamma}^{\nu}\Gamma^{\mu}
\end{array}\right).\label{eq:51}
\end{align}
Here we have defined the Lorentz transformation of quaternion four
dimensional spinor $\varphi(x)$ in $D=6$ space is as
\begin{align}
\varphi(x)\rightarrow & S^{-1}(\omega)\varphi(x')\label{eq:52}
\end{align}
where the operator $S(\omega)=exp(\frac{1}{2}\Sigma^{\mu\nu}\omega_{\mu\nu}$).
The $\omega_{\mu\nu}$is infinitesimal quaternion parameter and $\Sigma_{\mu\nu}$
are the generators of Lorentz group in $D=6$ space. So, the infinitesimal
change for spinor given by 
\begin{align}
\delta\varphi(x)= & \frac{1}{2}\Sigma^{\mu\nu}\omega_{\mu\nu}\varphi(x')\label{eq:53}
\end{align}
Thus the Lorenz generators $\Sigma_{\mu\nu}$ satisfy the following
relation of Lie algebra of $SO(1,5)$ i.e. 
\begin{align}
[\Sigma^{\mu\nu},\Sigma^{\rho\sigma}]=-i & (\eta^{\mu\rho}\Sigma^{\nu\sigma}+\eta^{\nu\sigma}\Sigma^{\mu\rho}-\eta^{\mu\sigma}\Sigma^{\nu\rho}-\eta^{\nu\rho}\Sigma^{\mu\sigma})\label{eq:54}
\end{align}
In order to describe the quaternions supersymmetry, it is customary
to extend the 4x4 representation of Lorentz group to $8\times8$ matrix
representation in terms of Weyl representation of $\Upsilon-$ matrices
i.e.

\begin{align}
\Upsilon_{W}^{\mu}= & \left(\begin{array}{cc}
0 & \gamma_{W}^{\mu}\\
\gamma{}_{W}^{\mu} & 0
\end{array}\right)\,\,\,(\mu=0\, to\,5)\label{eq:55}
\end{align}
which also satisfy the Clifford algebra relation of equation (\ref{eq:39}).
Similarly, the generators of Lorentz group for Weyl representation
are defined as

\begin{align}
\Xi^{\mu\nu}= & \frac{i}{4}(\Upsilon_{W}^{\mu}\Upsilon_{W}^{\nu}-\Upsilon_{W}^{\nu}\Upsilon_{W}^{\mu})=\left(\begin{array}{cc}
\Sigma^{\mu\nu} & 0\\
0 & \Sigma^{\mu\nu}
\end{array}\right).\label{eq:56}
\end{align}
It comes out to be fully reducible representation and reproduces the
Lorentz transformation for eight dimensional spinor in $D=6$ space
as
\begin{align}
\delta\Psi(x)= & \frac{1}{2}\Xi^{\mu\nu}\Omega_{\mu\nu}\Psi(x')\label{eq:57}
\end{align}
where the $\Omega_{\mu\nu}$ is infinitesimal antisymmetric quaternion
parameter and $\Psi(x)$ is eight dimensional Weyl spinor in $D=6$
space.

\section{Super-Poincare algebra in $D=6$ Space:}

The Lie algebra of Lorentz group $SO(1,5)$ in equation (\ref{eq:54})
may also be written by the generalization of equation (\ref{eq:17})
in the following expression i.e. 
\begin{align}
[M_{\mu\nu},M_{\rho\sigma}]=- & i(\eta_{\mu\rho}M_{\nu\sigma}+\eta_{\nu\sigma}M_{\nu\rho}-\eta_{\mu\sigma}M_{\nu\rho}-\eta_{\nu\rho}M_{\mu\sigma})\label{eq:58}
\end{align}
For the description of Super Poincare algebra in $D=6$ space, we
have defined the linear momentum operator $P_{\mu}\{\mu=0\, to\,5\}$
as the generators of translation symmetry in $D=6$ space. So, the
Poincare group in $D=6$ space is described in terms of commutation
rules between generators of homogeneous Lorentz group $M^{\mu\nu}$
and linear momentum operators $P^{\mu}$ in the following manner

\begin{align}
\left[P^{\mu},P^{\nu}\right]= & 0\nonumber \\
\left[M^{\mu\nu},P^{\rho}\right]= & i(\eta^{\nu\rho}P^{\mu}-\eta^{\mu\rho}P^{\nu})\nonumber \\
\left[M^{\mu\nu},M^{\rho\sigma}\right]= & -i(\eta^{\mu\rho}M^{\nu\sigma}+\eta^{\nu\sigma}M^{\mu\rho}-\eta^{\mu\sigma}M^{\nu\rho}-\eta^{\nu\rho}M^{\mu\sigma})\label{eq:59}
\end{align}
where $P_{\mu}$ has $6-$ generators while the $P_{\mu}$ and $M_{\mu\nu}$
spans the $15$ dimensional space of homogeneous Lorentz group in
$D=6$ space. As such, the Poincare algebra in $D=6$ space contains
$6$ translation, $10$ rotation and $5$ Lorentz boost generators.
The components of $M_{0k}\,(\forall k=1\, to\,5)$ are Lorentz boost
generators and $M_{ij}\,(\forall i,j=1\, to\,5)$ for angular momentum
operators in $D=6$ space. Here it should be noted that the angular
momentum in $D=6$ space is dyadic tensor $M_{ij}\,(\forall i,j=1\, to\,5)$.
Consequently the angular momentum and boost play different role for
$D=6$ space of $SO(1,5)$ Lorentz group. 

According to No-Go theorem of Coleman-Mandula \cite{key-3} ``The
most general Lie algebra of symmetries of S-matrix contain the energy-momentum
operator $P_{\mu}$, the Lorentz generator $M_{\mu\nu}$ and finite
numbers of Lorentz scalars $B_{l}$ which are the elements of compact
Lie algebra of internal symmetry''. But this restriction is avoided
by Haag-Lopuzansky-Sohnius \cite{key-3} by introducing commutators
in addition to the commutators in the symmetry group of S-matrix .The
introduction of anticommutator to commutator in symmetry Lie algebra
is called grading of the algebra and the whole Lie algebra is called
graded Lie algebra or superalgebra of S- matrix. So, the superalgebra
is the maximum extension of the Lie algebra of symmetry of S-matrix
that is possible. For the extension of Lie algebra of Poincare group
in $D=6$ space to superalgebra we describe $Z_{2}$ grading algebra
of this algebra such as

$L=L_{0}\oplus L_{1}$with properties

$L_{0}$ : Lie algebra of Poincare group $\{P_{\mu},M_{\mu\nu}\}$
in $8\times8$ matrix representation. 

\begin{singlespace}
$L_{1}$: Lie algebra of $Q_{a}\,[\forall a=1\, to\,8]$.
\end{singlespace}

\begin{spacing}{0.75}
$Q_{a}'s$ are eight dimensional, containing four dimensional two
component spinors
\begin{align}
Q_{a}= & \left(\begin{array}{c}
Q_{\alpha}\\
\bar{Q}_{\dot{\alpha}}
\end{array}\right)\:\:\;\quad(\forall\alpha=1,2,3,4)\label{eq:60}
\end{align}
 Defining the composition rule $\star$ in L such as
\begin{align}
\star:L\times L\rightarrow & L\nonumber \\
A\star B= & AB-(-1)^{g(L_{r})g(L_{s})}AB\:\:\;\quad(\forall i,j=0,1)\label{eq:61}
\end{align}
where $A\epsilon L_{r}\, B\epsilon L_{s}$and $A\star B\epsilon L_{r+s\, mod\,2}$.$g(L_{r})$,
$g(L_{s})$ are the order of grading for the sub -algebras $L_{r}$
and $L_{s}$ defined as
\end{spacing}

\begin{singlespace}
\begin{align}
g(L_{r})= & \begin{array}{cc}
0, & \:(for\, bosons)\\
1, & \quad\quad\quad\quad\quad\quad\quad\qquad(for\, fermions)\,\,(\forall r=0,1)
\end{array}\label{eq:62}
\end{align}
So, $g(L_{0})=0$ and $g(L_{1})=1$. Taking these considerations we
get the commutation relations as
\end{singlespace}
\begin{enumerate}
\item \noindent $:L_{0}\times L_{0}\rightarrow L_{0}$ whose commutation
rules are obtained in equation (\ref{eq:59}).
\item $:L_{0}\times L_{1}\rightarrow L_{1}$ which enables the following
commutation rules
\begin{align}
[P_{\mu},Q_{a}]= & 0\nonumber \\
{}[M_{\mu\nu},Q_{a}]= & -(\Xi_{\mu\nu})_{ab}Q_{b},\:\{a,b=1\, to\,8,\:\mu,\nu=0\, to\,5\}\label{eq:63}
\end{align}

\item :$L_{1}\times L_{1}\rightarrow L_{0}$ gives rise the following anti
commutation relations for spinors
\end{enumerate}
\begin{align}
\{Q_{a},Q_{b}\}\epsilon & L_{o}\label{eq:64}\\
\{Q_{a},\bar{Q}_{b}\}\epsilon & L_{o}.\label{eq:65}
\end{align}
 As such, the $L_{0}$ contain the generators of Poincare algebra
of $D=6$ space. So, there must be
\begin{align}
\{Q_{a},Q_{b}\}= & \alpha^{\mu}P_{\mu}+\beta^{\mu\nu}M_{\mu\nu}\label{eq:66}
\end{align}
where $\alpha^{\mu}=-2(\Upsilon^{\mu}C)_{ab}$ and $\beta^{\mu\nu}=(\Xi^{\mu\nu}C)_{ab}$
, $C$ is charge conjugation matrix and $\Xi_{\mu\nu}$ are the representations
of Lorentz algebra in $D=6$ space. However, by the generalized Jacobi
identity the second term $\beta^{\mu\nu}$ in equation (\ref{eq:66})
vanishes and hence, we get the anticommutator rule

\begin{align}
\{Q_{a},Q_{b}\}= & -2(\Upsilon^{\mu}C)_{ab}P_{\mu}.\label{eq:67}
\end{align}
Multiplying both side of the above equation by $C$ and imposing Majorana
condition $\left\{ (CQ)_{a}=-\bar{Q}_{a}\right\} $, we get

\begin{align}
\{Q_{a},\bar{Q}_{b}\}= & 2(\Upsilon^{\mu})_{ab}P_{\mu}\label{eq:68}
\end{align}
So we find out the representation of super-Poincare algebra in $D=6$
space as
\begin{align}
[M_{\mu\nu},M_{\rho\sigma}]= & -i(\eta_{\mu\rho}M_{\nu\sigma}-\eta_{\mu\sigma}M_{\nu\rho}-\eta_{\nu\rho}M_{\mu\sigma}+\eta_{\nu\sigma}M_{\mu\rho})\nonumber \\
{}[M_{\mu\nu},P_{\rho}]= & -i(\eta_{\mu\rho}P_{\nu}-\eta_{\nu\rho}P_{\mu})\nonumber \\
{}[P_{\mu},P_{\nu}]= & 0\nonumber \\
{}[P_{\mu},Q_{a}]= & 0\nonumber \\
{}[M_{\mu\nu},Q_{a}]= & -(\Xi_{\mu\nu})_{ab}Q_{b}\nonumber \\
\{Q_{a},\bar{Q}_{b}\}= & 2(\Upsilon^{\mu})_{ab}P_{\mu}\nonumber \\
\{Q_{a},Q_{b}\}= & -2(\Upsilon^{\mu}C)_{ab}P_{\mu}\nonumber \\
\{\bar{Q}_{a},\bar{Q}_{b}\}= & 2(C^{-1}\Upsilon^{\mu})_{ab}P_{\mu}\label{eq:69}
\end{align}
where the $\Xi_{\mu\nu}$ is the representation of Lorentz generator
acting on eight dimensional quaternionic spinor in $D=6$ space. $C$
is charge conjugation matrix in $D=6$ space which reduces to unity
in the case of Majorana representation. The Grassmann numbers $Q_{a}$
are invariant under the translation in space-time and thus commute
with the generators of momentum operators $P_{\mu}$,while transform
as spinors under Lorentz transformations $M_{\mu\nu}$. 

\newpage{}

\subsection*{Appendix I :The generators of homogeneous Lorentz group \textmd{$SO(1,5)$
}:}

Generators of Lorentz Boosts of section (3) in $SO(1,5)$ are described
as

\begin{align*}
N_{01}=\left(\begin{array}{cccccc}
0 & 1 & 0 & 0 & 0 & 0\\
1 & 0 & 0 & 0 & 0 & 0\\
0 & 0 & 0 & 0 & 0 & 0\\
0 & 0 & 0 & 0 & 0 & 0\\
0 & 0 & 0 & 0 & 0 & 0\\
0 & 0 & 0 & 0 & 0 & 0
\end{array}\right), & N_{02}=\left(\begin{array}{cccccc}
0 & 0 & 1 & 0 & 0 & 0\\
0 & 0 & 0 & 0 & 0 & 0\\
1 & 0 & 0 & 0 & 0 & 0\\
0 & 0 & 0 & 0 & 0 & 0\\
0 & 0 & 0 & 0 & 0 & 0\\
0 & 0 & 0 & 0 & 0 & 0
\end{array}\right),N_{03}=\left(\begin{array}{cccccc}
0 & 0 & 0 & 1 & 0 & 0\\
0 & 0 & 0 & 0 & 0 & 0\\
0 & 0 & 0 & 0 & 0 & 0\\
1 & 0 & 0 & 0 & 0 & 0\\
0 & 0 & 0 & 0 & 0 & 0\\
0 & 0 & 0 & 0 & 0 & 0
\end{array}\right)\\
N_{04}=\left(\begin{array}{cccccc}
0 & 0 & 0 & 0 & 1 & 0\\
0 & 0 & 0 & 0 & 0 & 0\\
0 & 0 & 0 & 0 & 0 & 0\\
0 & 0 & 0 & 0 & 0 & 0\\
1 & 0 & 0 & 0 & 0 & 0\\
0 & 0 & 0 & 0 & 0 & 0
\end{array}\right), & N_{05}=\left(\begin{array}{cccccc}
0 & 0 & 0 & 0 & 0 & 1\\
0 & 0 & 0 & 0 & 0 & 0\\
0 & 0 & 0 & 0 & 0 & 0\\
0 & 0 & 0 & 0 & 0 & 0\\
0 & 0 & 0 & 0 & 0 & 0\\
1 & 0 & 0 & 0 & 0 & 0
\end{array}\right)
\end{align*}
while the 10 generators of spacial rotations in $SO(1,5)$ are described
as:

\begin{align*}
L_{12}=\left(\begin{array}{cccccc}
0 & 0 & 0 & 0 & 0 & 0\\
0 & 0 & -1 & 0 & 0 & 0\\
0 & 1 & 0 & 0 & 0 & 0\\
0 & 0 & 0 & 0 & 0 & 0\\
0 & 0 & 0 & 0 & 0 & 0\\
0 & 0 & 0 & 0 & 0 & 0
\end{array}\right).L_{13}=\left(\begin{array}{cccccc}
0 & 0 & 0 & 0 & 0 & 0\\
0 & 0 & 0 & -1 & 0 & 0\\
0 & 0 & 0 & 0 & 0 & 0\\
0 & 1 & 0 & 0 & 0 & 0\\
0 & 0 & 0 & 0 & 0 & 0\\
0 & 0 & 0 & 0 & 0 & 0
\end{array}\right), & L_{14}=\left(\begin{array}{cccccc}
0 & 0 & 0 & 0 & 0 & 0\\
0 & 0 & 0 & 0 & -1 & 0\\
0 & 0 & 0 & 0 & 0 & 0\\
0 & 0 & 0 & 0 & 0 & 0\\
0 & 1 & 0 & 0 & 0 & 0\\
0 & 0 & 0 & 0 & 0 & 0
\end{array}\right)\\
L_{15}=\left(\begin{array}{cccccc}
0 & 0 & 0 & 0 & 0 & 0\\
0 & 0 & 0 & 0 & 0 & -1\\
0 & 0 & 0 & 0 & 0 & 0\\
0 & 0 & 0 & 0 & 0 & 0\\
0 & 0 & 0 & 0 & 0 & 0\\
0 & 1 & 0 & 0 & 0 & 0
\end{array}\right),L_{23}=\left(\begin{array}{cccccc}
0 & 0 & 0 & 0 & 0 & 0\\
0 & 0 & 0 & 0 & 0 & 0\\
0 & 0 & 0 & -1 & 0 & 0\\
0 & 0 & 1 & 0 & 0 & 0\\
0 & 0 & 0 & 0 & 0 & 0\\
0 & 0 & 0 & 0 & 0 & 0
\end{array}\right), & L_{24}=\left(\begin{array}{cccccc}
0 & 0 & 0 & 0 & 0 & 0\\
0 & 0 & 0 & 0 & 0 & 0\\
0 & 0 & 0 & 0 & -1 & 0\\
0 & 0 & 0 & 0 & 0 & 0\\
0 & 0 & 1 & 0 & 0 & 0\\
0 & 0 & 0 & 0 & 0 & 0
\end{array}\right)\\
L_{25}=\left(\begin{array}{cccccc}
0 & 0 & 0 & 0 & 0 & 0\\
0 & 0 & 0 & 0 & 0 & 0\\
0 & 0 & 0 & 0 & 0 & -1\\
0 & 0 & 0 & 0 & 0 & 0\\
0 & 0 & 0 & 0 & 0 & 0\\
0 & 0 & 1 & 0 & 0 & 0
\end{array}\right),L_{34}=\left(\begin{array}{cccccc}
0 & 0 & 0 & 0 & 0 & 0\\
0 & 0 & 0 & 0 & 0 & 0\\
0 & 0 & 0 & 0 & 0 & 0\\
0 & 0 & 0 & 0 & -1 & 0\\
0 & 0 & 0 & 1 & 0 & 0\\
0 & 0 & 0 & 0 & 0 & 0
\end{array}\right), & L_{35}=\left(\begin{array}{cccccc}
0 & 0 & 0 & 0 & 0 & 0\\
0 & 0 & 0 & 0 & 0 & 0\\
0 & 0 & 0 & 0 & 0 & 0\\
0 & 0 & 0 & 0 & 0 & -1\\
0 & 0 & 0 & 0 & 0 & 0\\
0 & 0 & 0 & 1 & 0 & 0
\end{array}\right)
\end{align*}

\begin{align*}
L_{45}= & \left(\begin{array}{cccccc}
0 & 0 & 0 & 0 & 0 & 0\\
0 & 0 & 0 & 0 & 0 & 0\\
0 & 0 & 0 & 0 & 0 & 0\\
0 & 0 & 0 & 0 & 0 & 0\\
0 & 0 & 0 & 0 & 0 & -1\\
0 & 0 & 0 & 0 & 1 & 0
\end{array}\right)
\end{align*}

\subsection*{Appendix-II :Dirac Representation:-}

The$8\times8$ fully reducible Dirac representation of $\Upsilon-$
matrrices in $D=6$ space is described as

\begin{align}
\Upsilon_{D}^{\mu}= & \left\{ \left(\begin{array}{cc}
\gamma_{D}^{0} & 0\\
0 & \gamma_{D}^{0}
\end{array}\right),\left(\begin{array}{cc}
\gamma_{D}^{1} & 0\\
0 & \gamma_{D}^{1}
\end{array}\right),\left(\begin{array}{cc}
\gamma_{D}^{2} & 0\\
0 & \gamma_{D}^{2}
\end{array}\right),\left(\begin{array}{cc}
\gamma_{D}^{3} & 0\\
0 & \gamma_{D}^{3}
\end{array}\right),\left(\begin{array}{cc}
\gamma_{D}^{4} & 0\\
0 & \gamma_{D}^{4}
\end{array}\right),\left(\begin{array}{cc}
\gamma_{D}^{5} & 0\\
0 & \gamma_{D}^{5}
\end{array}\right)\right\} \label{eq:70}
\end{align}
where $\gamma_{D}^{\mu}$ matrices are defined in equation (\ref{eq:39}).
For this Dirac representation the charge conjugation matrix is modified
as
\begin{align}
C_{D}=\left(\begin{array}{cc}
C_{4x4} & 0\\
0 & C_{4x4}
\end{array}\right),\, where\, & C_{4x4}=\left(\begin{array}{cccc}
0 & 0 & 0 & 1\\
0 & 0 & -1 & 0\\
0 & 1 & 0 & 0\\
-1 & 0 & 0 & 0
\end{array}\right)\label{eq:71}
\end{align}
this charge conjugation matrix satisfies the following properties 

(i) $C_{D}=-C_{D}^{\dagger}=-C_{D}^{-1}=C_{D}^{T}$ 

(ii)$C_{D}\Upsilon^{\mu}C_{D}^{-1}=-\Upsilon^{\mu T}$. 

The $(\Upsilon^{\mu}C_{D})$ comes out to be symmetric i.e. $(\Upsilon^{\mu}C_{D})^{T}=C_{D}^{T}\Upsilon^{\mu T}=\Upsilon^{\mu}C_{D}$.
Now post multiplying the equation (\ref{eq:67}) by $C_{D}$ given
in eq.(\ref{eq:71}) and applying Majorana condition $(C_{D}Q)_{a}=-\bar{Q}_{a}$
then we get 
\begin{align}
\{Q_{a},\bar{Q}_{b}\}= & 2(\Upsilon_{D}^{\mu})_{ab}P_{\mu}\label{eq:72}
\end{align}
which is the part of Super Poincare Algebra in $D=6$ space.

\subsection*{Appendix-III:Weyl Representation:-}

We may now identify 8x8 Weyl representation of $\Upsilon-$matrices
as 
\begin{align}
\Upsilon^{\mu}= & \left\{ \left(\begin{array}{cc}
0 & \gamma_{W}^{0}\\
\gamma_{W}^{0} & 0
\end{array}\right),\left(\begin{array}{cc}
0 & \gamma_{W}^{1}\\
\gamma_{W}^{1} & 0
\end{array}\right),\left(\begin{array}{cc}
0 & \gamma_{W}^{2}\\
\gamma_{W}^{2} & 0
\end{array}\right),\left(\begin{array}{cc}
0 & \gamma_{W}^{3}\\
\gamma_{W}^{3} & 0
\end{array}\right),\left(\begin{array}{cc}
0 & \gamma_{W}^{4}\\
\gamma_{W}^{4} & 0
\end{array}\right),\left(\begin{array}{cc}
0 & \gamma_{W}^{5}\\
\gamma_{W}^{5} & 0
\end{array}\right)\right\} \label{eq:73}
\end{align}
where $\gamma_{W}^{\mu}$ s are described in eq.(\ref{eq:49}). Here
the charge conjugation takes the following matrix representation i.e.
\begin{align}
C_{W}=\left(\begin{array}{cc}
C_{4x4} & 0\\
0 & C_{4x4}
\end{array}\right),\, where\, & C_{4x4}=\left(\begin{array}{cccc}
0 & 1 & 0 & 0\\
-1 & 0 & 0 & 0\\
0 & 0 & 0 & 1\\
0 & 0 & -1 & 0
\end{array}\right).\label{eq:74}
\end{align}
Charge conjugation matrix$(C_{W})$ follows the following properties:

(i) $C_{W}=-C_{W}^{\dagger}=-C_{W}^{-1}=C_{W}^{T}$ 

(ii)$C_{W}\Upsilon_{W}^{\mu}C_{W}^{-1}=-\Upsilon_{W}^{\mu T}$. The
$(\Upsilon_{W}^{\mu}C_{W})$ comes out to be symmetric $(\Upsilon_{W}^{\mu}C_{W})^{T}=C_{W}^{T}\Upsilon_{W}^{\mu T}=\Upsilon_{W}^{\mu}C_{W}$.
It is customary that the Parity $(P)$and Charge conjugation $(C)$
violate in Weyl representation. However the combined operation$CP$
is remains invariant in Weyl representation of Dirac equation. Now
post multiplying the equation (\ref{eq:67}) by $C_{W}$ given in
eq.(\ref{eq:74}) and applying Majorana condition $(C_{W}Q)_{a}=-\bar{Q}_{a}$
, we get

\begin{align}
\{Q_{a},\bar{Q}_{b}\}= & 2(\Upsilon_{W}^{\mu})_{ab}P_{\mu}\label{eq:75}\\
\left\{ \left(\begin{array}{c}
Q_{\alpha}\\
\bar{Q}_{\dot{\alpha}}
\end{array}\right)_{a},\left(\begin{array}{cc}
Q_{\beta} & \bar{Q}_{\dot{\beta}}\end{array}\right)_{b}\right\} = & 2\left(\begin{array}{cc}
0 & \left(\gamma_{W}^{\mu}\right)_{\alpha\dot{\beta}}\\
\left(\gamma_{W}^{\mu}\right)_{\dot{\alpha}\beta} & 0
\end{array}\right)_{ab}P_{\mu}\label{eq:76}
\end{align}
Likewise, we get the following relation for four component spinor

\begin{align}
\{Q_{\alpha},\bar{Q}_{\dot{\beta}}\}= & 2\left(\gamma_{W}^{\mu}\right)_{\alpha\dot{\beta}}P_{\mu},\,\{Q_{\alpha},Q_{\beta}\}=0\label{eq:77}\\
\{\bar{Q}_{\dot{\alpha}},\bar{Q}_{\dot{\beta}}\} & =0\label{eq:78}
\end{align}
$Q_{\alpha}$ is a four dimensional Weyl spinor such as
\begin{align}
Q_{\alpha}= & \left(\begin{array}{c}
Q_{l}\\
\bar{Q}_{\dot{l}}
\end{array}\right)\;\;,\,\bar{Q}_{\dot{\beta}}=\left(\begin{array}{cc}
Q_{m} & \bar{Q}_{\dot{m}}\end{array}\right)\:\:(l,m=1,2)\label{eq:79}
\end{align}
where $Q_{l}$ and $Q_{m}$ are two dimensional Weyl spinors. So by
substituting Q and $\bar{Q}$ from this equation we get
\begin{align}
\{Q_{l},\bar{Q}_{\dot{m}}\}= & 2\left(\Gamma^{\mu}\right)_{l\dot{m}}P_{\mu},\,\{Q_{l},Q_{m}\}=0\label{eq:80}\\
\{\bar{Q}_{\dot{l}},Q_{m}\}= & 2\left(\bar{\Gamma}^{\mu}\right)_{\dot{l}m}P_{\mu},\,\{\bar{Q}_{\dot{l}},\bar{Q}_{\dot{m}}\}=0\label{eq:81}
\end{align}
which is the part of Super Poincare Algebra in $D=6$ space.
\end{document}